\documentclass[
 reprint,
superscriptaddress,
bibnotes,
 amsmath,amssymb,
 aps,
pra,
]{revtex4-2}

\usepackage[a4paper,centering,hmargin=1.75cm,vmargin=2cm]{geometry} 
\usepackage{graphicx}
\usepackage{braket}
\usepackage{amsmath,bm,microtype}
\usepackage{mathrsfs}
\usepackage[dvipsnames]{xcolor} 

\usepackage{dcolumn}
\usepackage{colortbl}
\usepackage{siunitx}

\renewcommand{\vec}[1]{\bm{\mathrm{#1}}}

\DeclareFontFamily{U}{mathx}{\hyphenchar\font45}
\DeclareFontShape{U}{mathx}{m}{n}{
<-6> mathx5 <6-7> mathx6 <7-8> matha7
<8-9> mathx8 <9-10> mathx9
<10-12> mathx10 <12-> mathx12
}{}
\DeclareSymbolFont{mathx}{U}{mathx}{m}{n}
\DeclareMathSymbol{\bigtimes}{\mathop}{mathx}{"91}

\usepackage[colorlinks,allcolors=blue!50!black]{hyperref} 
\usepackage[all]{hypcap} 
\begin{document}

\title{Programmable generation of arbitrary  continuous-variable anharmonicities and nonlinear couplings}

\author{Teerawat Chalermpusitarak}
\email[Email: ]{teerawat.chalermpusitarak@sydney.edu.au}
\altaffiliation{These authors contributed equally to this work.}
\affiliation{School of Physics, University of Sydney, NSW 2006, Australia}

\author{Kai Schwennicke}
\email[Email: ]{kai.schwennicke@sydney.edu.au}
\altaffiliation{These authors contributed equally to this work.}
\affiliation{School of Chemistry, University of Sydney, NSW 2006, Australia}

\author{Ivan Kassal}
\affiliation{School of Chemistry, University of Sydney, NSW 2006, Australia}
\affiliation{Sydney Nano Institute, University of Sydney, NSW 2006, Australia}

\author{Ting Rei Tan}
\affiliation{School of Physics, University of Sydney, NSW 2006, Australia}
\affiliation{Sydney Nano Institute, University of Sydney, NSW 2006, Australia}


\begin{abstract}
Harmonic oscillators are promising continuous-variable (CV) quantum resources because their infinite-dimensional Hilbert spaces allow for resource-efficient quantum computing and simulation. To reach their full potential, CV platforms need to be able to efficiently implement non-Gaussian operations. Bosonic quantum-signal-processing schemes have emerged as attractive methods to construct arbitrary non-Gaussian operations; however, these schemes are restricted to single modes, i.e., the implementation of anharmonic potentials. Here, we introduce trigonometric-gate-implemented Fourier synthesis (TGIFS), a method for implementing arbitrary non-Gaussian operations applicable to both single- and multi-mode systems, allowing the generation of both anharmonicities and nonlinear multi-mode couplings. TGIFS synthesizes a target Hamiltonian by decomposing it into a Fourier series whose terms are implemented via bosonic quantum signal processing, which uses a discrete-variable (DV) system to induce a nonlinearity in the CV system. Our hybrid CV-DV protocol allows for the direct simulation of a broad range of CV phenomena (such as those in lattice gauge theory, chemical dynamics, and quantum chaos) and provides a richer toolbox for CV circuit compilation. 
\end{abstract}

\maketitle

Continuous-variable (CV) quantum platforms---such as microwave resonators~\cite{ofek2016extending,campagne2020quantum,Lescanne2020exponential,Yvonne_2021,Kudra2022robust,Cai2024protecting,Lachance2024autonomous,putterman2025preserving}, motional modes of trapped ions~\cite{toyoda2015hong,ortiz2017continuous,Fluhmann2019,Gan2020Hybrid,chen2023scalable,bazavan2024,Metzner2024twomode,jeon2024experimental,Hou2024,matsos2025universal,Li2025programmable}, and optical modes~\cite{bartlett2002universal,Yukawa2008experminetal,Asavanant2019generation,konno2024logical,larsen2025integrated,jhkz-84dz,Jai2025CVmultipartite}---are attractive resources for both quantum computation and simulation because a single CV system, such as a harmonic oscillator, has infinitely many orthogonal states, compared to the small number in typical discrete-variable (DV) systems, such as qubits. These infinite-dimensional Hilbert spaces enable quantum computation and simulation without discretization into DV registers~\cite{Loyd1999Quantum,Braunstein2005quantum,Navickas2025experimental,liu2025hybrid}, significantly reducing the hardware overhead~\cite{MacDonell2021Analog,crane2024hybrid}. For example, bosonic quantum error-correcting codes~\cite{Gottesman2001encoding,Terhal2020towards} use the infinite number of orthogonal states for redundancy. Furthermore, direct quantum simulation of physical CV systems is possible because their position and momentum quadratures can be mapped onto those of a CV device. This approach has led to the CV simulation of vibronic spectra~\cite{vibronic_spectra_estimation_Isaac, Macdonell2023predicting,Huh2015,Sparrow2018}, molecular dynamics~\cite{Gorman2018Engineering,MacDonell2021Analog,Valahu2023direct,Whitlow2023quantum,Wang2023ovservation,Navickas2025experimental,Visal2024trappedion,Fallas2025delocalized}, structured baths~\cite{Wang2024simulating,Sun2025}, and high-energy physics~\cite{Gerritsma2010,Gerritsma2011quantum, Gumpel2025analog,saner2025realtime}.

The capabilities of CV platforms can be expanded through the ability to implement anharmonicities and nonlinear couplings. Anharmonicities are Hamiltonian terms that are higher than quadratic with respect to a single oscillator's quadratures, while nonlinear couplings involve quadratures of multiple oscillators. Unitaries generated by these higher-than-quadratic Hamiltonians are called non-Gaussian unitary operations~\cite{Gaussian_quantum_information} and are essential to achieve CV platforms' full breadth of potential applications. For example, any universal gate set for CV quantum computation requires at least one non-Gaussian unitary, such as a Kerr-type or cubic interaction~\cite{Loyd1999Quantum,sefi2011decompose}.  Similarly, for CV quantum simulators to solve realistic problems in quantum chemistry or quantum field theory, they must incorporate both anharmonicities and nonlinear couplings~\cite{bender1969anharmonic,miller1983very,Silbey1980general,Munn1985theory,Brack1993quantum,Hamm2012vibraitonal,zhu2024extending,malpathak2025simulating,Li2025experimental,liu2025observation,Z2simulation}.


Engineering strong, arbitrary non-Gaussian unitary operations remains a central challenge for all CV platforms. For example, in photonic platforms, non-Gaussian operations require either materials with infeasibly high nonlinear susceptibilities or measurement-induced nonlinearities~\cite{shapiro2006single,knill2001scheme,kok2007linear} that impose significant hardware overhead~\cite{obrien2009photonic}. Specific anharmonicities and nonlinear couplings can be implemented in superconducting circuit quantum electrodynamics and trapped-ion devices with tailored hardware~\cite{Frattini20173wave,dealbornoz2024oscillatory,brown2011coupled,mielenz2016arrays}, 
but these approaches do not allow programming arbitrary non-Gaussian unitaries. Alternatively, anharmonicity and nonlinear couplings can be generated in trapped-ion platforms via resolved spin-oscillator transitions obtained from expanding the light-atom interaction with respect to the Lamb-Dicke parameter, $\eta$~\cite{Leibfried2002,Stobinska2011generation,Sameti2021strong,chen2023scalable}. However, higher-order terms in $\eta$ are weak because most systems work in the Lamb-Dicke regime, $\eta \ll 1$. 
This unfavourable $\eta$ scaling can be reduced by using non-commuting, linear CV-DV interactions to generate non-Gaussian interactions~\cite{sutherland,matsos2024,saner2024generating,Maverick2025}; however, these schemes do not provide a systematic compilation procedure and lack an analysis of how the error scales with the amount of resources. Therefore, it remains unclear which classes of Hamiltonians can be efficiently compiled within a target error.

To address these challenges, protocols based on bosonic quantum signal processing (QSP) have been developed to provide a systematic approach using a structured circuit to generate non-Gaussian unitaries on a single mode~\cite{QSPI,Lie2025Toward,hong2025oscillator,fong2025,singh2025nonabelian}. Bosonic QSP relies on a CV-DV architecture and uses a qubit as the DV system to impart anharmonicity onto the CV system. The anharmonicity arises from polynomial transformations of a signal operator, such as a displacement. These protocols have three desirable attributes. First, they are programmable using tunable CV-qubit interactions, rather than requiring customized hardware. Second, they can be systematically implemented using only single-qubit rotations and Gaussian CV operations, both of which are readily available on CV-DV devices. Third, bosonic QSP's structured circuits are provably efficient with a well-defined error scaling given a target polynomial transformation with one variable~\cite{LowQSP,AAQSP}. 

However, bosonic QSP has been limited to one mode so far. Extending it to multiple modes would be an instance of multivariate QSP (MQSP), for which a general theory is lacking due to difficulties in characterizing the class of multivariate polynomial transformations achievable through QSP-style ansatz circuits~\cite{Rossi2022multivariable,Mori2024comment,meth2023,Laneve2025multivariatev,iterateQSP,Rossi2025modularquantum,singh2025nonabelian}. It is therefore unclear if previous bosonic QSP methods developed for a single mode can accommodate multiple modes, especially when trying to synthesize nonlinear multi-mode couplings.  

Here, we develop a hybrid CV-DV scheme, trigonometric-gate-implemented Fourier synthesis (TGIFS), which incorporates bosonic-QSP as a subroutine to implement arbitrary non-Gaussian unitaries, including those with multi-mode couplings. TGIFS is systematic, fully analytical, and synthesizes non-Gaussian unitaries by expanding the corresponding non-quadratic Hamiltonians into multidimensional discrete Fourier series, whose components are implemented one-by-one using bosonic QSP. Unlike previous single-mode bosonic-QSP protocols that decompose the non-Gaussian unitary directly~\cite{park2024efficient,hong2025oscillator,Eberly}, our Hamiltonian-level decomposition naturally accommodates multiple modes through the use of trigonometric gates that are functions of linear combinations of the modes' quadratures, bypassing the complexities associated with characterizing the multivariate polynomial transformation. We demonstrate both programmability and systematic improvability of TGIFS and give examples showing that it can simulate the dynamics of a single anharmonic mode as well as of two nonlinearly coupled modes. We note that our approach has been implemented experimentally~\cite{cameron}, confirming the feasibility of the proposed approach.

\section{Theory}

TGIFS is a systematic method for generating arbitrary non-Gaussian unitaries that are functions of mutually commuting oscillator quadratures. This commutativity is essential, as it ensures the operators can be represented by the Fourier series expansions employed later in our derivation. We denote an arbitrary quadrature $Q_{n}^\theta=(a_ne^{-i\theta_n}+a_n^\dagger e^{i\theta_n})/\sqrt{2}$, where $a_n$ is the bosonic annihilation operator for mode $n$, and $\theta\in[0,2\pi)$. In particular, the position and momentum operators $X_n$ and $P_n$ are
\begin{equation}
    X_n = Q_{n}^{0} \quad \text{and} \quad P_n = Q_{n}^{\pi/2}.
\end{equation}
Our approach consists of decomposing the non-quadratic Hamiltonian, rather than the corresponding non-Gaussian unitary, into a Fourier series, whose components are then generated using bosonic QSP. This Hamiltonian-level---rather than unitary-level---decomposition enables the use of trigonometric gates that are functions of linear combinations of oscillator quadratures, allowing TGIFS to be naturally applicable to both the single- and multi-mode scenarios.

\subsection{Fourier decomposition of Hamiltonian}

We consider Hamiltonians that are functions of the commuting quadratures $\vec{Q}=(Q^{\theta_1}_{1},\cdots,Q^{\theta_n}_{n})$. We will synthesize Hamiltonians $V(\vec{Q})$ that can be written as a multi-dimensional Fourier series
\begin{align}\label{eq:targetHam}
    V(\vec{Q}) = \sum_{\vec{m}} C_{\vec{m}} \exp\left(i\sum_{n=1}^{N}  m_n k_nQ_n^{\theta_n}\right),
\end{align}
where $\sum_{\vec{m}}=\sum_{m_1=-\infty}^{\infty}\sum_{m_2=-\infty}^{\infty}\cdots\sum_{m_N=-\infty}^{\infty}$, $m_n$ is an integer that refers to the $m$th Fourier component of the $n$th coordinate, and $k_n=2\pi/L_n$, where the $L_n$ are chosen so that the region of interest falls within $[-L_n/2,L_n/2]$ (outside of which the Fourier expansion is periodic). The size of the domain can be enlarged at the cost of including more Fourier terms to maintain accuracy. The Fourier coefficients $C_{\vec{m}}$ are
\begin{align}
    C_{\vec{m}} = \frac{1}{|\mathcal{V}|}\int_\mathcal{V}\text{d}^{N}\vec{q}\,V(\vec{q}) \exp\left(-i\sum_{n=1}^{N}m_nk_n q_n^{\theta_n}\right),
\end{align}
where the integral is over the domain $\mathcal{V}=\bigtimes_{n=1}^{N}[-L_n/2,L_n/2]$ with volume  $|\mathcal{V}|= \prod_{n=1}^{N}L_n$ and $Q_n^{\theta_n}\ket{q_n^{\theta_n}} = q_n^{\theta_n}\ket{q_n^{\theta_n}}$. Alternatively, using the fact that $V(\vec{Q})$ is Hermitian, we can represent it as a sum of sines and cosines of the quadratures 
\begin{multline}\label{eq:surfacepotential}
    V(\vec{Q})= \sum_{\vec{\mu}=\vec{0}}^\infty \big(A_{\vec{\mu}}\cos(\vec{\mu}\cdot\vec{Q}) + B_{\vec{\mu}}\sin(\vec{\mu}\cdot\vec{Q})\big),
\end{multline}
where $\vec{\mu}=(\mu_1,\ldots,\mu_N)=(m_1k_1,\ldots,m_Nk_N)$, and $A_{\vec{0}}=C_{\vec{0}}$, $A_{\vec{\mu}\neq\vec{0}}=2\text{Re}\,C_{\vec{m}}$, and $B_{\vec{\mu}}=-2\text{Im}\, C_{\vec{m}}$ are the Fourier coefficients. 

The time evolution of the system under $V(\vec{Q})$ for a time step $\Delta t$ is given by the unitary operator 
\begin{align}
    U(\Delta t) &= \exp(-iV(\vec{Q})\Delta t),\\
    &= \prod_{\vec{\mu}=\vec{0}}^{\infty} T_c(\vec{\mu},\Lambda_{\vec{\mu}}^c) T_s(\vec{\mu},\Lambda^s_{\vec{\mu}}) \label{eq:expansion},
\end{align}
where we set $\hbar=1$ and defined the \textit{trigonometric gates}
\begin{align}
    T_c(\vec{\mu},\Lambda^c_{\vec{\mu}}) &= \exp(-i\Lambda_{\vec{\mu}}^{c}\cos(\vec{\mu}\cdot\vec{Q})),\\
    T_s(\vec{\mu},\Lambda^s_{\vec{\mu}}) &= \exp(-i\Lambda_{\vec{\mu}}^s\sin(\vec{\mu}\cdot\vec{Q})),
\end{align}
with $\Lambda_{\vec{\mu}}^c=A_{\vec{\mu}}\Delta t$ and $\Lambda_{\vec{\mu}}^s=B_{\vec{\mu}}\Delta t$. The expansion in Eq.~\ref{eq:expansion} is exact because the quadratures on different modes commute. Therefore, we can simulate the dynamics under the Hamiltonian $V(\vec{Q})$ given the ability to implement the two families of trigonometric gates.

\begin{figure*}[!ht]
\begin{centering}
\includegraphics[width=\linewidth]{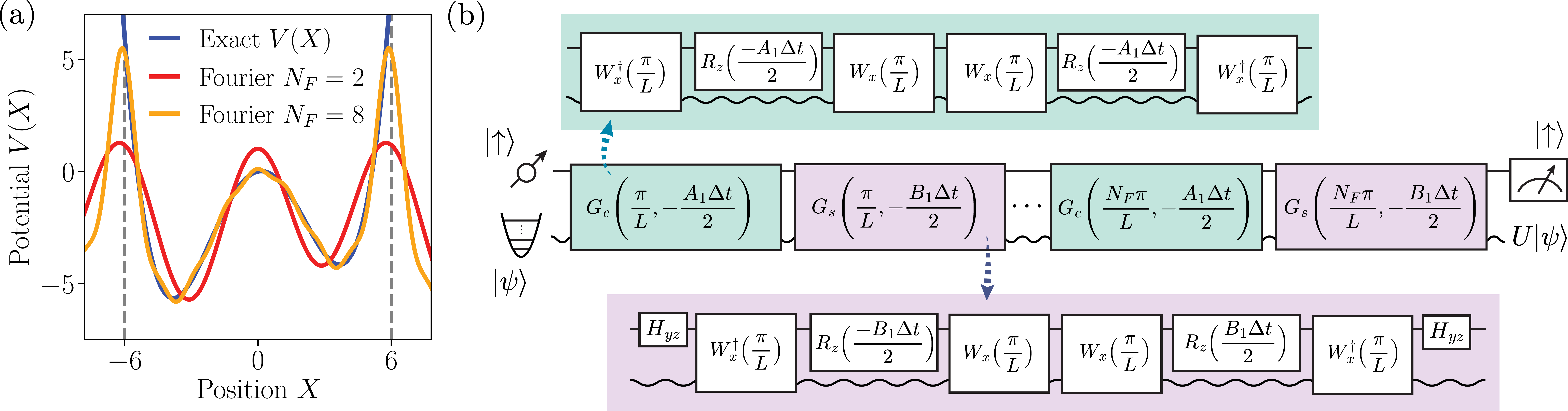}
\end{centering}
\vspace{-2mm}
\caption{TGIFS decomposition of an anharmonic Hamiltonian into CV-DV trigonometric gates. \textbf{(a)} A target Hamiltonian, the double-well potential $V(X) = 0.05X^4-0.7X^2+0.2X$ (blue), is approximated by the Fourier series $\sum_{m=1}^{N_F} A_{m}\cos(mX)+B_{m}\sin(mX)$ truncated at the 2nd or 8th orders. Dashed lines depict the domain of consideration ($L=12$). 
\textbf{(b)} Middle row: CV-DV circuit for generating arbitrary single-mode potentials. Each term in the Fourier series of $V(X)$ is implemented using the corresponding qubit-dependent trigonometric gate $G_{c,s}$. Postselection at the end of the circuit removes the dependence on the qubit, leaving the boson-only unitary $U\approx e^{-iV(X)\Delta t}$. Top and bottom rows: Bosonic QSP sequences for the gates $G_{c,s}$.\label{fig:circuit}} 
\end{figure*}

\subsection{Trigonometric gates via bosonic QSP}\label{subsection:generation of Gs}

To generate the nonlinear gates $T_c$ and $T_s$, we introduce one ancilla qubit, which allows us to use bosonic QSP sequences consisting of single-qubit rotations and bosonic operations conditioned on the qubit state. To do so, we adapt to the multi-mode setting a bosonic QSP sequence designed for single-mode Fourier synthesis~\cite{QSPI,Lie2025Toward,hong2025oscillator}. We construct $T_c$ and $T_s$ in two steps, by first synthesizing qubit-dependent versions of the gates and then eliminating the qubit degree of freedom.

Our bosonic QSP sequence is given by the unitary
\begin{align}
    U_\text{QSP}(\vec{\phi}) = R_z(\phi_d)\prod_{j=0 }^{d-1}W_x(\vec{\kappa}) R_z(\phi_j),\label{eq:bosonic-qsp}
\end{align}
where $R_z(\phi)=e^{i\phi\sigma_z}$ is a single-qubit rotation about the $z$-axis and $W_x(\vec{\kappa}) = e^{i \sigma_x  \vec{\kappa}\cdot\vec{Q}}$ is a conditional multi-mode displacement with $\vec{\kappa}=(\kappa_1,\ldots,\kappa_N)$ being a vector containing the amount of displacement for each mode. 
The multi-mode operator $W_x(\vec{\kappa})$ can be decomposed as
\begin{align}
    W_x(\vec{\kappa}) = \prod_{j=1}^{N} e^{i\sigma_x\kappa_j Q_j^{\theta_j}},
\end{align}
meaning that it can be implemented using single-mode state-dependent displacements, which are ubiquitous in CV-DV platforms, including trapped ions~\cite{wineland1998experimental} and superconducting circuits~\cite{blais2004cavity,touzard2019gated,eickbusch2022fast}. In trapped-ion architectures, the number of available motional modes increases with the number of ions, and a single qubit can be coupled to each spectrally resolved mode via individual state-dependent displacements generated by laser–ion interactions~\cite{Lee2005phase,Friis2018observation}. In superconducting circuits, multi-mode resonators coupled to a qubit provide similar access to multiple bosonic modes within the same device~\cite{Sundaresan2015beyond,kollar2019hyperbolic}.

The gates $T_c$ and $T_s$ can be generated using the QSP sequence in Eq.~\ref{eq:bosonic-qsp} with $d=2$ and angles
\begin{align*}
    \phi_0 = \pi/2, \quad \phi_1 = -\pi/2+\vartheta, \quad \phi_2 = 0, \quad\text{for $\vartheta\in [0,2\pi)$}.
\end{align*}
This sequence results in the unitary operator
\begin{align}
    \mathcal{U}(\vartheta,\vec{\kappa}) &= W_x(\vec{\kappa})e^{i\vartheta\sigma_z}e^{-i\frac{\pi}{2}\sigma_z}W_x(\vec{\kappa})e^{i\frac{\pi}{2}\sigma_z},\\
    &= W_x(\vec{\kappa}) e^{i\vartheta\sigma_z} W_x^\dagger(\vec{\kappa}),\\
    &= e^{i\vartheta(\sigma_z\cos(2\vec{\kappa}\cdot\vec{Q})+\sigma_y\sin(2\vec{\kappa}\cdot\vec{Q}))} \label{eq:QSP1},
\end{align}
Since $T_c$ and $T_s$ are generated by the operators $\text{sin}(\vec{\kappa}\cdot\vec{Q})$ and $\text{cos}(\vec{\kappa}\cdot\vec{Q})$, we need to separate those two contributions in Eq.~\eqref{eq:QSP1}. To do so, we use the Lie-Trotter formula \cite{trotter1959product} and define two qubit-dependent unitaries
\begin{align}
G_c(\vec{\kappa},\vartheta) &= \mathcal{U}(\vartheta,\vec{\kappa}) \mathcal{U}(\vartheta,-\vec{\kappa})\nonumber\\
   &\approx \exp\big(2i\vartheta\sigma_z\cos( 2\vec{\kappa}\cdot\vec{Q})\big),\label{eq:spin-dependent cosine}\\
   G_s(\vec{\kappa},\vartheta) &= H_{yz}\mathcal{U}(\vartheta,\vec{\kappa})\mathcal{U}(-\vartheta,-\vec{\kappa})H_{yz}\nonumber\\
   &\approx \exp\big(2i\vartheta\sigma_z\sin( 2\vec{\kappa}\cdot\vec{Q})\big),
    \label{eq:spin-dependent sine}
\end{align}
where $H_{yz} = (\sigma_y+\sigma_z)/2$ is the Hadamard gate used to convert $\sigma_y$ to $\sigma_z$. These approximations are valid for $\vartheta\ll 1$.
The gates $G_{c,s}$ are qubit-dependent versions of $T_{c,s}$, provided that we set $\vartheta=-\Lambda_{\vec{\mu}}^{c,s}/2$ and $\kappa_n=\mu_n/2$.

The qubit-independent $T_{c,s}$ can be obtained by applying $G_{c,s}$ to an initial state $\ket{\uparrow,\psi_{\text{osc}}}$. Because the qubit state $\ket{\uparrow}$ is unaffected by $G_{c,s}$, we can deterministically post-select on it to obtain
\begin{align}
   T_{c,s}(\vec{\mu},\Lambda^{c,s}_{\vec{\mu}}) |\psi_{\text{osc}}\rangle &= \bra{\uparrow}G_{c,s}\left(\frac{\vec{\mu}}{2},-\frac{\Lambda^{c,s}_{\vec{\mu}}}{2}\right)\ket{\uparrow}|\psi_{\text{osc}}\rangle. \label{eq:elementary-trig-gates}
\end{align}
As an example, Fig.~\ref{fig:circuit} shows the implementation of TGIFS for simulating a double-well Hamiltonian.

\section{Examples}
\label{sec:examples}
We present two simulations to illustrate the range of Hamiltonians that TGIFS can synthesize: a one-mode anharmonic potential and two-mode nonlinear coupling. In both examples, simulation accuracy increases with the number of Fourier components, showing that TGIFS can be systematically improved.

\begin{figure*}[!ht]
\begin{centering}
\includegraphics[width=\linewidth]{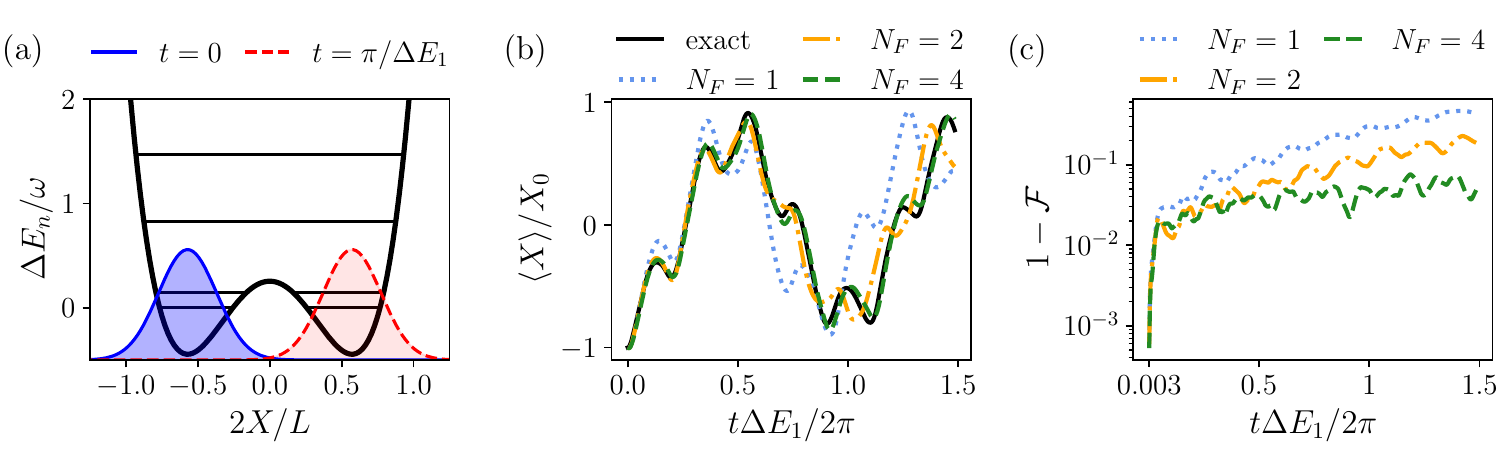}
\end{centering}
\vspace{-5mm}
\caption{Simulating quantum tunneling in an anharmonic potential using TGIFS. 
\textbf{(a)}~Quartic double-well potential of Eq.~\ref{eq:double-well-hamiltonian}, with $\xi_1/\omega=8\xi_0/\omega=0.35$, where $\Delta E_{n}$ is the energy difference between the $n$th eigenstate and the ground state. The system is initially in the coherent state $\ket{\alpha=-X_0/\sqrt{2}}$, centered at the left well. Although the energy of the initial state is below the barrier, the system can tunnel into the right well, with time scale inversely proportional to $\Delta E_{1}$. 
\textbf{(b)}~Expected position~$\langle X\rangle$ of the oscillator as the CV-DV system is propagated using Eq.~\ref{eq:tunnelling-unitary} with different numbers of Fourier components $N_F$. The time step is $\Delta t=20\pi/(\omega r)$ with Trotter number $r=500$. As $N_F$ increases, so does the agreement with the exact dynamics.  \textbf{(c)} The infidelity $1-\mathcal{F}$ of the final state with respect to the exact result decreases with $N_F$, especially at larger times where the accumulated influence of anharmonicity is greater.  \label{fig:double-well}} 
\end{figure*}

\subsection{Anharmonic dynamics: Double-well potential}

We simulate the dynamics of a particle in the anharmonic quartic-double-well potential
\begin{equation}
    H_\text{dw}=\frac{\omega}{2}P^2+\xi_1 X^4-\xi_0 X^2. \label{eq:double-well-hamiltonian}
\end{equation} 
Double-well Hamiltonians are ubiquitous across physics, chemistry, and biology, particularly for modeling tunneling processes, with applications from quantum field theory~\cite{shuryak1988toward} to molecular isomerization~\cite{cribb1982comparison,Vardi2000role} and proton transfer in biology~\cite{godbeer2015modelling,slocombe2022open}.

To simulate tunneling under $H_\text{dw}$, we express the anharmonic component of the potential as a discrete Fourier series with $N_F$ terms,  
\begin{align}
H_{\text{dw}}^{F}&= H_0+\sum_{m=1}^{N_F} A_m\cos( mkX), \\
H_0&=\frac{\omega}{2}(P^2+X^2),\end{align} 
where $A_m$ are the Fourier coefficients of $V(X)=\xi_1 X^4-(\xi_0+\omega/2)X^2$. Since we are interested in the tunneling dynamics from one well to the other, we choose $L=7X_0/2$ (where $X_0$ is the location of the right-well minimum), which ensures that most of the dynamics will not leak outside the domain $[-L/2,L/2]$. Because $V(X)$ is even, there are no sine terms in the Fourier expansion. Additionally, the $m=0$ term can be ignored because it only leads to a global phase in the dynamics.

To simulate the dynamics under $H_{\text{dw}}^{F}$ for a total simulation time $T$ with time step $\Delta t=T/r$, we Trotterize it to first order. For a single time step,
\begin{align}
U_{\text{trot}}(\Delta t)&=e^{-iH_0\Delta t}U_{\text{dw}}^F(\Delta t), \label{eq:tunnelling-unitary}\\
U_{\text{dw}}^F(\Delta t)&=\prod_{m=1}^{N_F}G_c(mk/2,A_m\Delta t/2).
 \label{eq:dynamics-trotter-scheme}
\end{align}  
Here, the harmonic component is implemented through the free evolution of the CV system under $H_0$ and the individual Fourier components that represent the anharmonicity are implemented using our $G_c$ gates.

Upon initializing the CV-DV system in the state $\ket{\uparrow,\psi_{\text{osc}}}$ and then postselecting on $\ket{\uparrow}$ at the completion of the simulation, the CV system is in the state
\begin{equation}
    \langle\uparrow|(U_{\text{dw}}^F(\Delta t))^r|\uparrow\rangle|\psi_{\text{osc}}\rangle\approx e^{-iH_{\text{dw}}^{F}r\Delta t}\ket{\psi_{\text{osc}}},\label{eq:post-select-dymanics}
\end{equation}
i.e., the bosonic system follows the dynamics under $H_{\text{dw}}^{F}$. 

This simulation approach is systematically improvable. As $N_F$ and $r$ increase, the dynamics converges to the exact dynamics under $H_{\text{dw}}$. As shown in Fig.~\ref{fig:double-well}, the infidelity $1-\mathcal{F}$ decreases with $N_F$, where 
\begin{equation}
\mathcal{F}=|\langle{\psi_\text{osc}} |\exp(iH_{\text{dw}}T)\langle \uparrow|(U_{\text{dw}}^F(\Delta t))^r|\uparrow\rangle |\psi_\text{osc}\rangle|^2.
\end{equation}

\subsection{Two-mode nonlinear coupling}

\begin{figure*}
    \centering
    \includegraphics[width=\linewidth]{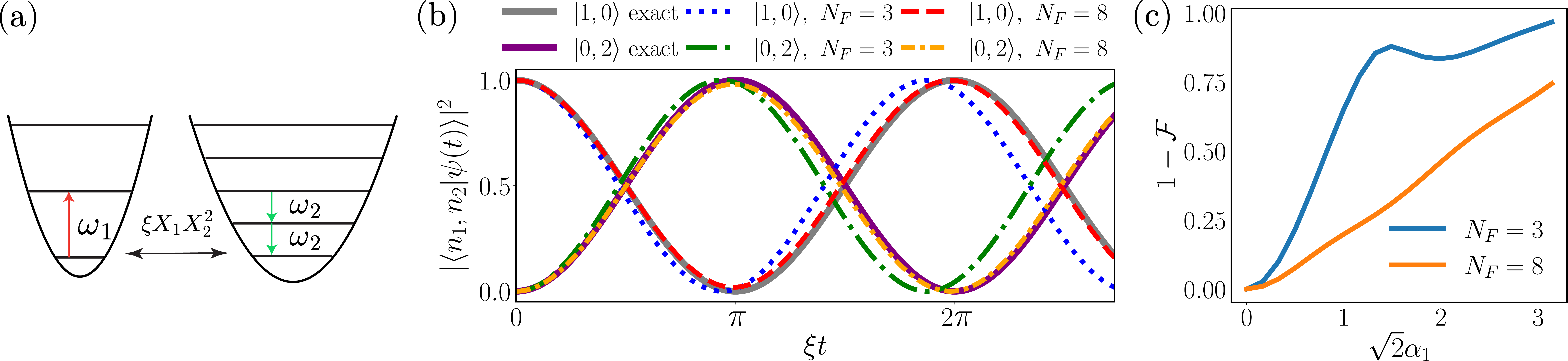}
    \caption{Simulating the dynamics of two nonlinearly coupled oscillators using TGIFS. \textbf{(a)} The coupled modes, with $\omega_2 = \omega_1/2$ and $\xi\ll \omega_1$. \textbf{(b)} Time evolution of the populations of the multi-mode Fock states $\ket{1,0}$ and $\ket{0,2}$, both exact and using Fourier expansions with $N_F$ of 3 or 8. \textbf{(c)} The infidelity $1-\mathcal{F}$ of quantum states $(U_{\rm trot}(\Delta t))^r\ket{\alpha_1,\alpha_2=0}$ with respect to the exact state $\exp(-iH_{\text{nl}}r\Delta t)\ket{\alpha_1,\alpha_2=0}$ improves as $N_F$ increases. For these simulations, $\xi\Delta t=1.715\times 10^{-3}$ and $r = 2500$.
    \label{fig:2modeexample}}
\end{figure*}

As another example, we generate nonlinear coupling between two harmonic oscillators and simulate their coupled dynamics. In our model (Fig.~\ref{fig:2modeexample}a), the oscillator frequencies are $\omega_1$ and $\omega_2=\omega_1/2$, and the nonlinear-coupling Hamiltonian is 
\begin{align}
    V_\text{nc} = \xi X_1X_2^2.
\end{align}
For $\xi\ll \omega_1$, $V_\text{nc}$ has been used to model both the nonlinear coupling that leads to Fermi resonance~\cite{Nosenko2018anharmonicity} as well as the anharmonic quantum battery, whose charging dynamics saturates the quantum speed limit \cite{Quantumbattery}.

The dynamics of the coupled system is described by the total Hamiltonian
\begin{align}\label{eq:two-coupled-HO-ham}
    H_{\rm nl} &= H_0^{(1)}+H_0^{(2)} + V_\text{nc}\\
    &= \frac{\omega_1}{2}(X_1^2+P_1^2)+\frac{\omega_2}{2}(X_2^2+P_2^2) + \xi X_1X_2^2,
\end{align}
where the first two terms describe the free evolution of the two oscillators.

To implement $V_\text{nc}$, we expand it as a two-dimensional Fourier series within the domain $(X_1,X_2)\in[-L_1/2,L_1/2]\times[-L_2/2,L_2/2]$,
\begin{align}
    V_\text{nc} &\approx V_\text{nc}^{F}=\sum_{m_1=1}^{N_F}\sum_{m_2=1}^{N_F} B_{m_1,m_2}\sin(m_1 k_1X_1+m_2k_2 X_2),
\end{align}
where $B_{m_1,m_2}$ are the Fourier coefficients and the cosine terms are excluded because $V_\text{nc}$ is odd under $(X_1,X_2)\rightarrow(-X_1,-X_2)$. We then use our trigonometric gates to construct the unitary
\begin{align}
    U_{\text{nc}}^F(\Delta t) &= \prod_{m_1,m_2=1}^{N_F}G_s\Big(\Big(\tfrac{m_1k_1}{2},\tfrac{m_2k_2}{2}\Big),-\frac{B_{m_1k_1,m_2k_2}\Delta t}{2}\Big),\\
    &\approx \exp\Big(-i\sigma_z \sum_{\vec{\mu}} B_{\vec{\mu}}\Delta t\sin(\mu_1 X_1+\mu_2 X_2)\Big).
\end{align}

We can now Trotterize the dynamics under $H_\text{nl}$, with the unitary operator of one time step being
\begin{align}
    U_{\rm trot}(\Delta t) &= e^{-i(H_0^{(1)}+H_0^{(2)})\Delta t}U_{\rm nc}^F(\Delta t)\\ 
    & \approx \exp\big(-i(H_0^{(1)}+H_0^{(2)}+\sigma_zV^{F}_\text{nc})\Delta t\big).
\end{align}
The final state of the oscillators at time $T=r\Delta t$ can be simulated by applying $U_{\rm trot}(\Delta t)$ $r$ times to the initial state $\ket{\uparrow}\ket{\psi_{\text{osc}}(0)}$, followed by postselection on $\ket{\uparrow}$, i.e., 
\begin{align}
    \ket{\psi_{\text{osc}}(T)} &= \langle\uparrow|(U_{\rm trot}(\Delta t))^r |\uparrow\rangle|\psi_{\text{osc}}(0)\rangle\label{eq:trotterized_dynamics}\\
    &\approx e^{-iH_{\rm nl} r\Delta t}|\psi_{\text{osc}}(0)\rangle.
\end{align}

Fig.~\ref{fig:2modeexample} demonstrates the accuracy of TGIFS by simulating the dynamics of the two coupled modes initialized in the multi-mode Fock state $\ket{1,0}$. The dynamics in Fig.~\ref{fig:2modeexample}b shows population transfer between the states $\ket{1,0}$ and $\ket{0,2}$, which is only possible due to the nonlinear coupling. The accuracy of the Trotterized dynamics improves as $N_F$ increases. 

In addition, we demonstrate the accuracy of TGIFS for various initial states whose wavefunctions lie within the domain given by $L_1=L_2=2\pi$. To do so, we repeat our simulations on initial states that are tensor products of coherent states, $\ket{\alpha_1,\alpha_2=0}$. Fig.~\ref{fig:2modeexample}c shows that TGIFS is particularly accurate for states close to the origin (small $\alpha_1$) and that its accuracy can be improved with increasing $N_F$ at all $\alpha_1$.

\section{Discussion}

TGIFS is a programmable CV-DV protocol that systematically generates both arbitrary anharmonicities and arbitrary nonlinear multi-mode couplings. It synthesizes the non-Gaussian unitaries by implementing the terms in the Fourier series of their non-quadratic Hamiltonians via the trigonometric gates, rather than implementing the Fourier expansion of the unitaries~\cite{park2024efficient,hong2025oscillator}. This strategy is crucial to the applicability of TGIFS to multi-mode scenarios because it gives analytic expressions for all control parameters, such as the angles of the single-qubit rotations and the amounts of displacement. Therefore, it avoids optimization-based phase-finding algorithms used in existing unitary-based bosonic QSP~\cite{park2024efficient,hong2025oscillator}, which are inapplicable to the multi-mode case~\cite{Rossi2022multivariable,Mori2024comment,Laneve2025multivariatev,ito2025polynomial}.

The accuracy of the non-Gaussian operations generated by TGIFS is systematically improvable by increasing $r$, $N_F$, and $L_n$. By increasing $r$ for a fixed $T$, we reduce $\Delta t$, thereby decreasing the Trotter error associated with generating each trigonometric gate (Eqs.~\ref{eq:spin-dependent cosine}--\ref{eq:spin-dependent sine}). Increasing $N_F$ improves the Fourier-series expansion of the Hamiltonian within the domain specified by $L_n$. To extend the dynamics outside of a particular domain, the Fourier expansion can be recalculated for a larger $L_n$, although doing so may require more Fourier components to reach the same accuracy.

To discuss the scalability of implementing nonlinear multi-mode couplings, we distinguish the total number of modes, $N$, from the largest number of modes that appear in a coupling term, $M$. The number of Fourier components, and thus the number of trigonometric gates required, scales as $\mathcal{O}(N^M N_F^M)$. For a fixed $M$ (e.g., in local Hamiltonians), this scaling is polynomial in the total number of modes. However, for $M=N$ (all-to-all coupling), the number of trigonometric gates grows exponentially with the number of modes, which restricts the scalability of TGIFS for implementing highly multi-mode interactions. Nevertheless, TGIFS remains scalable in the more common scenarios where many modes are present but only a few appear in individual coupling terms, as arise in quantum signatures of chaos~\cite{Brack1993quantum}, nonlinear vibronic-coupling models~\cite{Hamm2012vibraitonal}, and nonlinear-coupling models for Fermi resonances ~\cite{Nosenko2018anharmonicity}.

TGIFS can substantially decrease the time required to implement non-Gaussian operations on a trapped-ion device compared to generating the same operations using interactions that are higher-order in the Lamb-Dicke parameter $\eta$. For a single $d$th-order coupling term in the Lamb-Dicke expansion that contains $M$ modes, the interaction strength scales as $\mathcal{O}(\eta^d)$, requiring a gate time of $\mathcal{O}(1/\eta^d)$. In contrast, the timescales of the single-qubit rotations and the conditional displacements scale as $\mathcal{O}(1)$ and $\mathcal{O}(1/\eta)$, respectively. Because each trigonometric gate requires four conditional displacements, the total time to generate the same $d$th-order non-Gaussian operation containing $M$ modes goes as $\mathcal{O}(4N_F^M/\eta)$. Therefore, TGIFS achieves a net speedup compared to the native non-Gaussian implementation when the number of Fourier components meets the requirement $N_F<\mathcal{O}\big(\eta^{-(d-1)/M}\big)$. For example, for a typical $\eta=0.01$, a speedup for $d=4$ and $M=1$ (e.g., a Kerr gate) is possible for $N_F\lesssim10^6$. Therefore, the speedup can be multiple orders of magnitude, especially for single-mode anharmonicities. 

TGIFS provides access to a wide range of non-Gaussian single- and multi-mode gates that could have broad applications in quantum computing and simulation. In quantum computing, our expanded gate set could lead to more compact CV-DV circuit decompositions through quantum-compilation techniques for identifying shorter gate sequences. In quantum simulation, TGIFS supports the construction of non-Gaussian operations for simulating complex quantum dynamics. Given that the required low-order Gaussian operations are already available with high fidelity on existing CV–DV devices, we expect these advances to be experimentally accessible on near-term hybrid quantum platforms.

\acknowledgments{We were supported by the Australian Research Council (FT220100359, FT230100653), the United States Office of Naval Research Global (N62909-24-1-2083), the Wellcome Leap Quantum for Bio program, Lockheed Martin, the Sydney Horizon Fellowship (TRT), and H.\ and A.\ Harley. We thank Henry Nourse, Mohammad Nobakht, Cameron McGarry, Christophe Valahu, Frank Scuccimarra, Maverick Millican, Vanessa C.~Olaya-Agudelo, Vassili G.~Matsos, Liam Flew, and Prachi Nagpal for valuable discussions.}

\bibliographystyle{apsrev4-2}
\bibliography{bib}

\end{document}